# First-principles study of metal-graphene edge contact for ballistic Josephson junction

Yeonghun Lee,[1] Jeongwoon Hwang,[1,3] Fan Zhang,[2] and Kyeongjae Cho[1]

[1]Department of Materials Science and Engineering, University of Texas at Dallas, Richardson, Texas 75080, USA

[2]Department of Physics, University of Texas at Dallas, Richardson, Texas 75080, USA

[3]Department of Physics Education, Chonnam National University, Gwangju 61186, South Korea

**Abstract**

Edge-contacted superconductor-graphene-superconductor Josephson junctions have been utilized to realize topological superconductivity, and have shown superconducting signatures in the quantum Hall regime. We perform first-principles calculations to interpret electronic couplings at the superconductor-graphene edge contacts by investigating various aspects in hybridization of molybdenum $d$ orbitals and graphene $\pi$ orbitals. We also reveal that interfacial oxygen defects play an important role in determining the doping type of graphene near the interface.

# I. Introduction

Inducing superconductivity into quantum Hall systems has been a long-standing challenge in condensed matter and materials physics. Recently, this has reemerged as a major theme in the quest for nonabelian Majorana fermions and parafermions, which can naturally appear if topological edge states become superconducting [1–4]. Finding these nonabelian quasiparticles is extremely rewarding, because they offer a fundamentally different promise to store and manipulate quantum information in a nonlocal and decoherence-free fashion.

Yet, the magnetic field for realizing quantum Hall effect breaks time-reversal symmetry, which is essential for conventional superconductors. For the coexistence of superconductivity and quantum Hall effect, it is necessary that the upper critical field of the superconductor must be higher than the magnetic field threshold for the emergence of the quantum Hall effect. This challenge can be overcome by using high-quality h-BN-encapsulated graphene. h-BN encapsulation can efficiently reduce atomic defects and produce ultrahigh-mobility graphene [5] exhibiting integer (fractional) quantum Hall effect at a magnetic field as low as 1 T (5 T) [6,7]. Such field strengths are much lower than the upper critical fields of several type-II superconductors.

Another requirement is to achieve a sufficiently transparent superconductor-graphene interface that can yield an efficient superconducting proximity effect. In other words, suppressed normal scattering at the interface effectively produces the Cooper pair in the superconductor and a retroreflected electron or hole in the graphene region through Andreev reflection. Only in this way, the quantum Hall edge states can be made superconducting or mediate the supercurrent of Cooper pairs. On top of earlier efforts on the realizations of superconducting graphene [8–14], most recently, both the Josephson effect and crossed Andreev reflection have been observed in the integer quantum Hall regime [7,15,16]. Such compelling evidence marks a significant step toward topological superconductivity.

Particularly, to observe the supercurrent in the quantum Hall regime, Amet *et al.* [7] fabricated MoRe-(hBN-encapsulated graphene)-MoRe ballistic Josephson junctions, based on the prior success demonstrated by Calado *et al.* [15] The MoRe alloy is a type II superconductor with a high upper critical field of 8 T at 4 K [7,15]. Importantly, to attain transparent contacts, the devices designed for the experiments [7,11–16] utilized the one-dimensional interfacial edge contact between superconductor and graphene. In addition, this edge-contact geometry enables to separate layer assembly and contact metallization processes so that polymer contamination can be avoided [17]. While experimental progress based on such edge-contact Josephson junctions has been actively reported, there have been very few theoretical studies on superconductor-graphene interfaces. Specifically, it has not yet been thoroughly investigated with realistic interface atomic and electronic structures why the MoRe-graphene edge contact can enable the superb proximity effect.

In this work, we perform first-principles calculations based on density functional theory (DFT) [18,19] to shed light on the electronic structures of superconductor-graphene edge contacts. We also directly estimate the contact transparency through the transmission coefficient, derived from a combined method of DFT simulation and nonequilibrium Green's function formalism (DFT-NEGF) [20,21]. To reduce complexity, we adopt the body-centered cubic (bcc) Mo to model the MoRe alloy and focus on the intrinsic characteristics of Mo-graphene edge contact. This approximation can be justified as follows. The superconducting MoRe alloy used in experiments [7,11,22] is a 50/50 wt% mixture of Mo and Re. According to phase diagrams [23], it is expected that the MoRe alloy does not form a single phase but forms a mixed phase of two different phases—bcc Mo and intermediate χ phase—at room temperature. DFT modeling studies [24,25] have shown that the surface energy of Mo(110) is lower than that of Re(0001). Consequently, Mo is most likely to be exposed at the surface and form a contact with graphene

The interface can be characterized by Mo $d$ orbitals and graphene $\pi$ orbitals, both of which mainly distribute electronic states around the Fermi level. The $\pi$ electrons are loosely bound to carbon atoms and contribute to supercurrent transport. By systematically investigating the orbital-decomposed band structures, projected densities of states (PDOS), band gaps, local potentials, etc., we examine the chemical bonding and strong $d$-$\pi$ orbital hybridization at the Mo-graphene edge contact. It is known that they are absent in the side-contact geometry for graphene [26]. This difference would imply that the edge contact is advantageous for attaining an atomically smooth, transparent Josephson junction with strong electron coupling and low contact resistance that should enable the observed superb proximity effect. For the MoRe-graphene interface, we revisit the above consensus pertaining to advantageous edge contact. Finally, we examine the effects of interfacial oxygen defects, which could be possibly induced during $O_2$ plasma etching, and explain well the $n$-type doping of interfacial graphene in the experiments [7,11,15]. For this analysis, we take into account the work function difference between graphene and superconductor, which eventually affects the transmission at the contact [14].

## II. Method

To investigate the electronic structures, we carry out the first-principles calculations based on the DFT [18,19] method using the Vienna Ab initio Simulation Package (VASP) [27,28]. The exchange-correlation energy functional is given by the Perdew-Burke-Ernzerhof (PBE) functional [29] in the generalized gradient approximation (GGA). The pseudopotential is given by the projector-augmented wave (PAW) method [30,31]. The kinetic energy cutoff for the plane-wave basis set is 400 eV, and a uniform Γ-centered $k$-point grid of 1×15×5 is adopted for the Brillouin zone sampling of rectangular supercells. Spin polarized calculations are performed.

The DFT-NEGF transport simulations are performed with the TranSIESTA code [32]. The GGA-PBE functional is adopted as above. The norm-conserving pseudopotentials are constructed with the Troullier-Martins scheme [33]. As a linear combination of atomic orbital (LCAO) basis set, the single-ζ-polarized basis set (SZP) is employed along with 0.05 Ry of the pseudoatomic orbital (PAO) energy shift. [34] The real space mesh cutoff is set to 150 Ry, and spin polarized calculations are performed as well. We employ the supercell structures that are previously optimized with VASP. In TranSIESTA inputs, the supercells need to be rotated to make the transport direction, which is originally the $x$ direction as shown in Fig. 1, point in the $z$ direction. We also shorten the graphene sheet because the transport simulation effectively deals with infinite electrodes outside of a simulation domain. With the modified supercells (see Fig. S2 in the Supplemental Material), a Monkhorst-Pack $k$-point grid of 3×21×1 is adopted for self-consistent calculations, and that of 3×1001×1 is adopted to calculate the transmission coefficient.

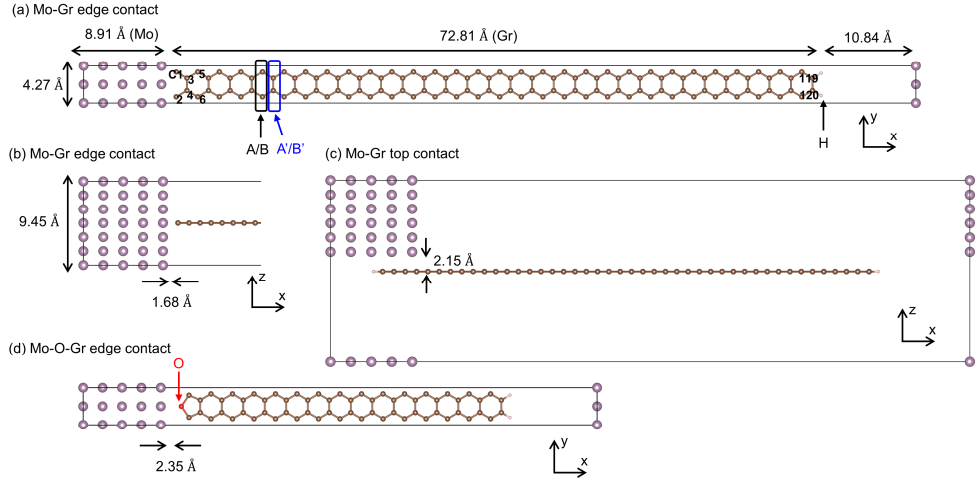

Fig. 1. (a) Top view and (b) side view of the supercell structure of Mo-graphene edge contact. At the interface, Mo(110) contacts one armchair graphene edge. The optimized distance between Mo and the graphene edge is 1.68 Å. Dangling bonds at the other armchair edge are passivated by hydrogens. Each carbon atom is numbered from 1 to 120. Indices A/B and A'/B' for dimer lines are adopted to distinguish carbon groups, each of which consists of graphene sublattices A and B. (c) Supercell structure of Mo-graphene top contact with optimized spacing, 2.15 Å. (d) Supercell structure of Mo-graphene edge contact with an interfacial oxygen defect. The oxygen defect is 2.35 Å away from the Mo surface.

To model the one-dimensional interfacial edge contact between Mo and graphene, we adopt the (110) surface of bcc Mo and the armchair edge of graphene (Fig. 1). Although the hBN-encapsulation changes the dielectric environment around graphene, we expect that the effect is weak because it does not fully cover the contact region, and the hBN is not explicitly included in the modeling study. For bcc metals, the (110) surface has the lowest surface energy among low index (100), (110), and (111) surfaces. The lattice mismatch between Mo and the armchair edge of graphene is only 4 % facilitating the interface modeling with small lattice mismatch. We manually optimize the model interface structure of the Mo-graphene edge contact. Also, we have checked further relaxation of the manually optimized interfacial structure, where the maximum force is smaller than 0.02 eV/Å. In the additional relaxation step, the largest displacement of each atom was 0.08 Å. Since this change is negligible, we adopted the interface model without further relaxation after manually adjusting the Mo-carbon distance. For comparison of edge and top contacts, we also built a top contact structure with a Mo(100)-graphene interface [Fig. 1(c)]. In addition to the ideal edge contact models, we have developed an oxygen-terminated graphene edge model by inserting an oxygen defect at the interface to analyze the doping type of the interfacial graphene. Fig. 1(d) plots the supercell atomic structure for this Mo-graphene edge contact with an interfacial oxygen defect. To obtain this model structure, the graphene edge with oxygen is relaxed while maintaining a pentagonal structure [35]. Subsequently, the distance between the metal surface and the oxygen atom is adjusted. This modeling sequence is consistent with the fabrication process flow, in which metal leads are deposited after an oxygen plasma etch [17]. See Fig. S1 in the Supplemental Material for the detailed relaxation procedures for the ideal interfaces and the oxygen-inserted one.

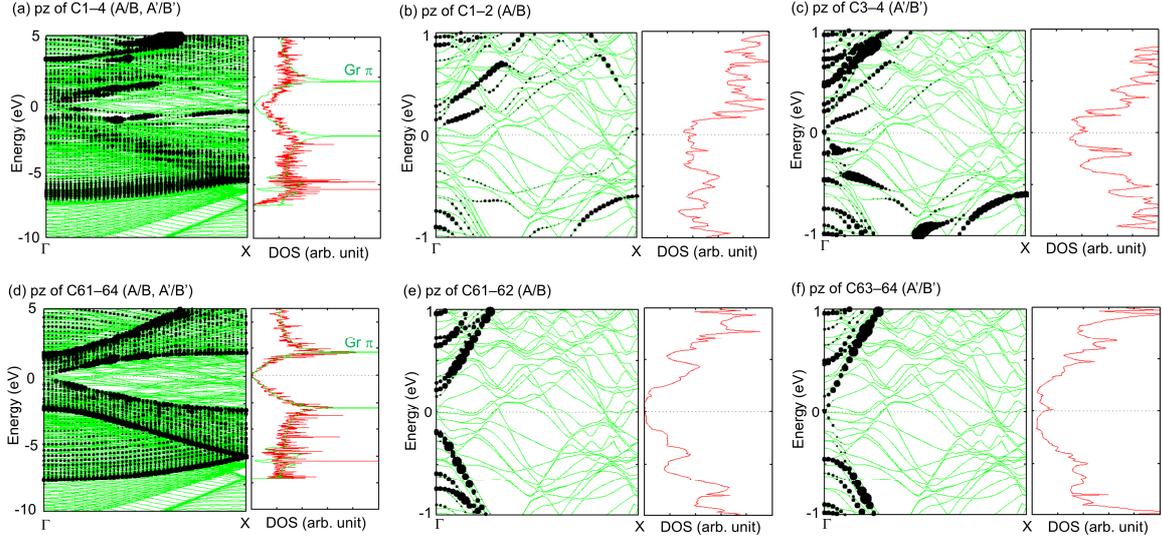

Fig. 2. Orbital-decomposed band structures (left panels) and PDOS (right panels) for carbon $p_z$ orbitals near the interface [(a) C1–4, (b) C1–2, and (c) C3–4] and inside the graphene [(d) C61–64, (e) C61–62, and (f) C63–64]. The Fermi energy is set to zero. The size of each symbol corresponds to the intensity of each orbital component. There is no bandgap inside the graphene. Interfacial electronic structures are strongly modified by hybridization with Mo $d$ orbitals.

## III. Results and discussion

### 1. Electronic structure analysis

To examine the electronic couplings between Mo $d$ orbitals and graphene $\pi$ orbitals, we evaluate orbital-decomposed band structures and PDOS for carbon $p_z$ orbitals, which are associated with graphene $\pi$ orbitals (Fig. 2) [36]. By comparing with carbon $p_z$ orbitals of an infinite graphene sheet, we find strong orbital hybridization of Mo $d$ orbitals and carbon $p_z$ orbitals near the interface. Particularly, the Mo $d$ states distribute within an energy range from –5 to 5 eV (Fig. S3 in the Supplemental Material), and thus carbon $p_z$ orbitals are strongly modified and hybridized in this energy range [Fig. 2(a)]. This chemical hybridization can give rise to transparent contacts for edge contact structures. In the interior of graphene, the carbon $p_z$ orbitals remain intact as they form $\pi$ orbitals similar to those in the infinite graphene sheet [Fig. 2(d)]. Since the band bending of graphene occurs within a limited distance, the Dirac point of interior graphene remains at the Fermi level.

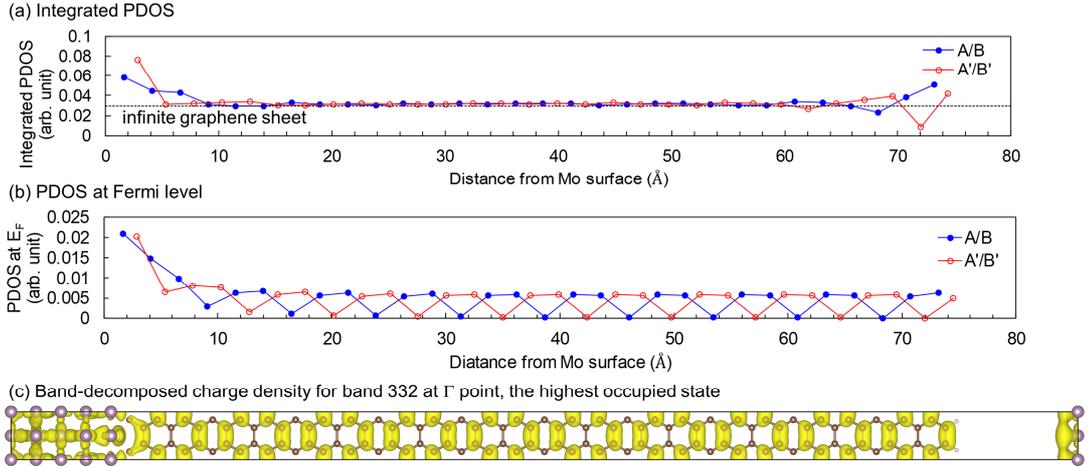

Fig. 3. (a) Integrated PDOS as a functional of distance from the Mo surface. The PDOS for carbon $p_z$ orbitals is integrated over energy from -1 to 1 eV. The integrated PDOS for an infinite graphene sheet is 0.0294 (arbitrary unit). (b) PDOS at the Fermi level oscillates with distance from the Mo surface. (c) Band-decomposed charge density for band 332 at the Γ point. The band-decomposed charge density corresponds to the probability density of the eigenstate.

The A and B carbons in each A/B dimer (also, the A' and B' carbons in each A'/B' dimer) have the same electronic structure due to a mirror symmetry of the supercell. In contrast, A/B and A'/B' dimers exhibit different electronic structures, as shown in Fig. 2(b), 2(c), 2(e), and 2(f). To quantitatively evaluate the electronic coupling of Mo and graphene, we integrate the PDOS of carbon $p_z$ orbitals over the energy range from –1 to 1 eV [37], which is plotted as a function of the distance from the Mo surface. Fig. 3(a) indicates that the coupling effect of Mo $d$ states penetrates about 10 Å into graphene. We also find that the integrated PDOS of A'/B' near the interface is greater than that of A/B, although the A'/B' dimer is farther away from the interface than the A/B dimer. However, the large integrated PDOS for A'/B' more rapidly decays in the graphene region. This can be explained via the complex band structure analysis [35,38]. The decay length is short when the imaginary component of the complex wave vector is large, and the imaginary component is positively correlated to the energy difference between the corresponding state and the conduction band minimum or valance band maximum; i.e., the larger the energy difference, the shorter the decay length. In the orbital-decomposed band structures for A'/B' [left panel of Fig. 2(c)], there is a large number of carbon $p_z$ states near the X point, and the energy gap is large at the *X* point. Based on the complex band structure analysis, it is consistent that the integrated PDOS for A'/B' more rapidly decay.

The right panels of Fig. 2(e) and 2(f) show that the PDOS of C63–64 exhibits a peak at the Fermi level whereas there is no peak for that of C61–62. For further analysis, we extract the PDOS for the peak at the Fermi level as a function of the distance from the Mo surface [Fig. 3(b)]. Mixing with Mo $d$ orbitals, the PDOS at the Fermi level is modified and becomes larger near the interface, within 10 Å. Even in the interior of the graphene, the peak exists and oscillates along the graphene interior. This oscillation can also be observed in the band-decomposed charge density [35]. The state near the Fermi level at the Γ point is almost threefold degenerate and consists of bands 332, 333, and 334. Fig. 3(c) plots the band-decomposed charge density for band 332, which is the valence band maximum. Evidently, the probability density of the eigenstate directly corresponds to the oscillation of the PDOS at the Fermi level.

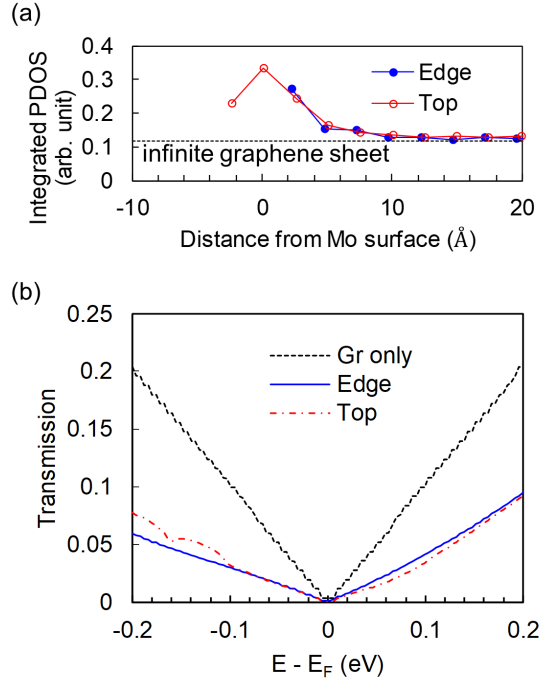

Fig. 4. (a) Integrated PDOS as a function of distance from the Mo surface. Each point is a summation over integrated PDOS for A, B, A', and B' carbon atoms. The position of the Mo surface is set to 0. (b) Transmission spectra derived from the DFT-NEGF. These are transmission coefficients summed over all the conducting channels for the given unit cells. Degrees of hybridization for edge and top contacts are nearly the same.

It is known that, for metal-graphene interfaces, edge contact geometry is advantageous because a short bond distance at the interface facilitates strong hybridization, whereas distant bond and weak hybridization are likely to form with top contact structures [17,26]. To address superior performance of the edge contact, Matsuda *et al*. employed Ti, Pd, Pt, Cu, and Au as a metal lead [26]. For top contact geometry, Pd, Pt, Cu, and Au exhibit weak hybridization with graphene. This weak interface interaction is because $d$ orbitals of those elements are fully occupied, and antibonding states can be filled with electrons. This $d$-orbital configuration destabilizes the metal-carbon bond, resulting in weak hybridization with graphene $\pi$ electrons in the top contact interface [39]. In this case, edge contact exhibits stronger hybridization with graphene $\sigma$-electron dangling bonds and lower contact resistance than top contact. On the other hand, Ti $d$ orbitals are partially filled, so that antibonding states are empty. Although a degree of the hybridization is not enough to obtain transparency comparable to edge contact, the Ti-graphene top contact shows strong hybridization. Interestingly, Mo-graphene top contact also exhibits strong hybridization because Mo $d$ orbitals are partially occupied in $4d^5$ configuration (see Fig. S2 in the Supplemental Material), and transparency is comparable to edge contact. Fig.

4(a) shows that a similar degree of hybridization is observed from electronic structure modifications in edge and top contacts. Furthermore, transmission spectra derived by using a combined DFT-NEGF method result in the fact that contact transmission is nearly the same for those contact geometries [Fig. 4(b)]. Therefore, it is not always true that edge contact is accompanied by better transparency than top contact. We could attain transparent interface in Mo-graphene top contact. Since Re exhibits strong interaction due to its partially filled $d$ orbitals ($5d^5$) [40,41], transparent contact could also form at the Re-graphene interface as well as the Mo-graphene interface. Therefore, we expect to find similar discussion for the MoRe-graphene top contact.

Graphene nanoribbons with armchair edges can be classified into three distinct types: $3n$, $3n+1$, and $3n+2$, where $n$ is a positive integer [35,42,43]. The numeric labels correspond to the number of dimer lines across the ribbon width. Since examining the detailed differences between the three types is beyond the scope of this paper, we have chosen to use the $3n$ type ($n$=10, 20) in our study. Nevertheless, it is worthwhile pointing out that the three types have different band gaps, where the $3n+2$ type exhibits the smallest band gap for hydrogen-passivated armchair edges on both sides [42,43].

As for our model structures, one armchair graphene edge contacts the Mo surface, and the other edge is passivated by hydrogen. In this case, instead of the $3n+2$ type, the $3n$ type exhibits the smallest band gap, which almost vanishes [Fig. 2(f)]. See Fig. S4 in the Supplemental Material for the result that the band gaps for the other two types are greater than 0.1 eV. Ryou *et al.* [35] have reported that, compared with the hydrogen-passivated one, edge-functionalized graphene nanoribbons have a different width dependence of band gaps. Similarly, the edge-contacted graphene nanoribbons are also different from the hydrogen-passivated one. In our model structure, the $p_z$ orbitals of interfacial carbons are strongly hybridized with the Mo $d$ orbitals and largely deviate from their original $\pi$-orbital properties. This results in the greatly reduced probability density at the interfacial carbon atoms [Fig. 3(c)]. At the hydrogen-passivated edge, the probability density of $p_z$ orbitals vanishes at the passivating hydrogen sites rather than the edge carbon sites. Thus, we can conclude that the edge-contacted interface makes the width effectively shorter in terms of the width dependence of band gaps; precisely, our edge-contacted $3n$ type corresponds to the $3n–1$ type for the typical hydrogen-passivated armchair graphene nanoribbons.

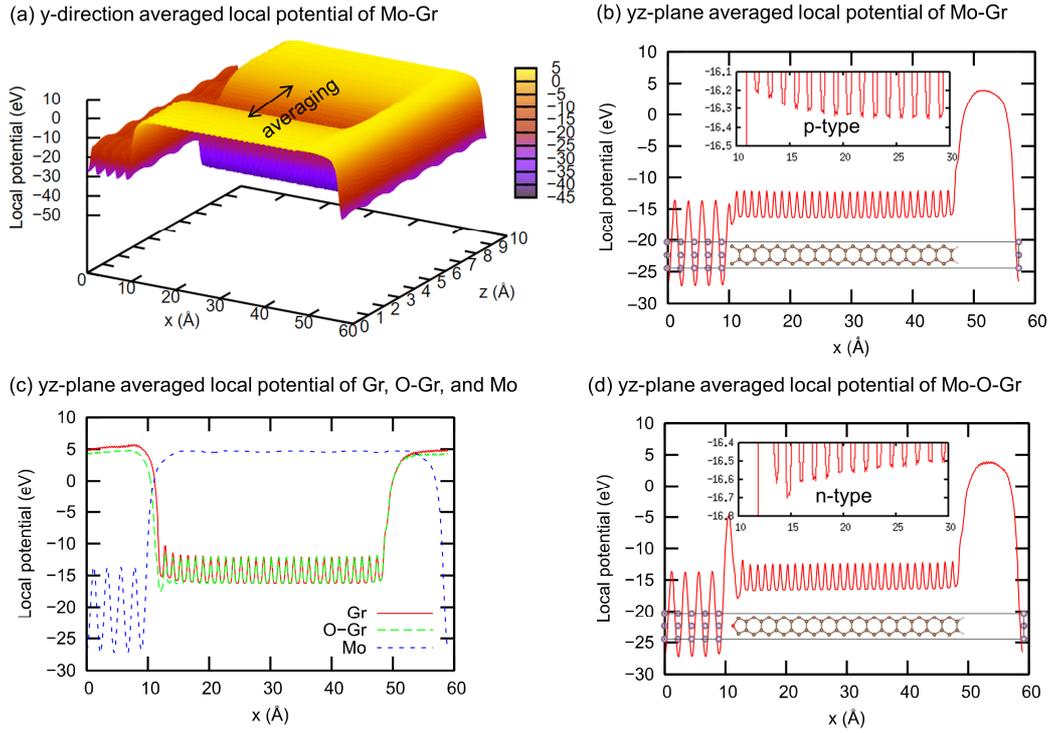

Fig. 5. (a) *y*-direction averaged local potential of Mo-graphene edge contact. (b) *yz*-plane averaged local potential of Mo-graphene edge contact. (c) *yz*-plane averaged local potentials of graphene, O-graphene, and Mo segments. (d) *yz*-plane averaged local potential of Mo-O-graphene edge contact. For a *yz*-plane average, a local potential is averaged over *z* within the interlayer distance of graphite. The Fermi level is set to zero. Insets of (b,d) are zoomed-in plots to distinguish different doping types of graphene at the interface.

## 2. Doping type of graphene at interface

Experiments have shown that *n*-type graphene is formed at the MoRe-graphene edge contact [11,15]. To investigate doping types for the edge contact structure, first we average the local potential along the *y* direction [Fig. 5(a)]. The local potential here includes the exchange correlation potential, as well as the ionic and Hartree potentials. To eliminate the contribution of the vacuum region, we take an additional average of the local potential within the interlayer distance of graphite, i.e., 3.4 Å [44]. Now we can recognize the doping type of graphene at the interface by using the *yz*-plane averaged local potential [Fig. 5(b)–(d)]. Contrary to the experimental result, *p*-type graphene is formed at the ideal Mo-graphene edge contact (inset of Fig. 5(b)). This result can be expected from the fact that the calculated work functions of Mo(110) and graphene are 4.74 and 4.26 eV, respectively. Although the lead in our model structure is Mo instead of the MoRe alloy, we can deduce that graphene at an ideal MoRe-graphene edge contact is also likely to be *p*-type, as the work function of Re is 4.95 eV [45]. Since the work functions of both Mo and Re are much greater than that of graphene, the work function of the MoRe alloy should be greater than that of graphene. Hence, there must be a nontrivial reason for the discrepancy between the computational and experimental results, since neither the Schottky-Mott rule nor the Fermi level pinning [46] can explain the difference.

To address this difference, it is realistic to include possible defects at the interface, which would be introduced in experimental interface formation. Among all types of defects, oxygen is most likely to be incorporated during the $O_2$ plasma etching process, which has been previously introduced to show an increase in transmission for a metal-graphene edge contact [17,44]. During the fabrication process, oxygen can passivate the graphene edge before the deposition of the MoRe alloy, and eventually interfacial oxygen can be placed between MoRe and graphene [Fig. 1(d)]. Our local potential calculation shows that this oxygen-incorporated structure gives rise to *n*-type graphene at the interface [Fig. 5(d)]. Fig. 5(c) indicates that the obtained *n*-type graphene is attributed to oxygen's high electronegativity, which enables electrons to transfer from graphene to oxygen. After oxygen passivation, the work function at the graphene edge becomes higher due to the presence of an electric field created by the electron transfer. A zigzag edge graphene nanoribbon has also shown the same change in doping type with respect to the interfacial oxygen [44]. We vary the distance between the Mo surface and oxygen atoms within the range from 1.5 to 2.7 Å, and the interface graphene is still of *n* type (see Fig. S5 in the Supplemental Material). It is likely that the *n*-type preference of interface graphene is a general feature for MoRe-graphene edge contact fabricated by using $O_2$ plasma etching.

One may be concerned that the interfacial barrier shown in Fig. 5(d) may not be consistent with the low contact resistance reported in previous research works [7,11,16,17]. We argue that the interfacial oxygen may not be an issue for achieving low contact resistance. First, the barrier height is lower than the Fermi level. Next, according to results by Gao and Guo [44], the transmission coefficient would not be drastically degraded even if the distance between metal and oxygen was 3 Å even though a vacuum gap that much wider at the interface might induce even higher potential barrier. Third, Wang *et al.* [17] have already shown that an oxygen defect can improve the contact resistance at the interface of chromium and the zigzag edge of graphene. Finally, there is a metastable interface structure with a much smaller distance between the oxygen atom and the Mo surface when the graphene layer is shifted along the *z* direction (see Fig. S1(f) in the Supplemental Material). For these reasons, an oxygen defect can play a role in forming *n*-type graphene without degrading transparency.

## IV. Conclusion

In summary, we have carried out first-principles study on the Mo-graphene edge contact for the experimentally established ballistic MoRe-graphene-MoRe Josephson junctions. PDOS analysis reveals that the Mo *d* and carbon $p_z$ orbitals strongly hybridize at the interface, which should account for the observed transparent Josephson junctions and superb proximity effects. We have further analyzed the evolutions of the integrated PDOS and observed the strong couplings within 10 Å of graphene. The PDOS analysis and transport simulation have indicated that, unlike other transition metals, Mo-graphene top contact could also exhibit transparent contact, attributed to the strong hybridization. In addition, we have shown that the edge-contacted graphene nanoribbon is effectively shorter than the hydrogen-passivated one in terms of width dependence of the band gap. Finally, as demonstrated above, the interfacial oxygen could be the origin for the *n*-type graphene observed in experiment, although *p*-type graphene is expected at the ideal MoRe-graphene interface. These results have established a quantitative theoretical basis to systematically interpret edge-contacted superconductor-graphene Josephson junctions and may pave the way for designing better devices for realizing topological superconductivity.

**Acknowledgments**

This work was supported by grants from the U.S. Army Research Laboratory's Army Research Office (ARO) (Grant No. W911NF-17-1-0369 and No. W911NF-18-1-0416). We acknowledge the Texas Advanced Computing Center (TACC) for providing supercomputing resources. We thank Dr. Francois Amet at Appalachian State University, Dr. Gleb Finkelstein at Duke University, Dr. James Williams at the University of Maryland, Dr. Marc Ulrich at U.S. Army Research Office, and Dr. Wei Pan at Sandia National Laboratories for valuable discussions.

**Supplemental material for "First-principles study of metal-graphene edge contact for ballistic Josephson junction"**

Yeonghun Lee,[1] Jeongwoon Hwang,[1,3] Fan Zhang,[2] and Kyeongjae Cho[1]

[1]Department of Materials Science and Engineering, University of Texas at Dallas, Richardson, TX 75080, USA

[2]Department of Physics, University of Texas at Dallas, Richardson, TX 75080, USA

[3]Department of Physics Education, Chonnam National University, Gwangju 61186, South Korea

## 1. Structure relaxation

The structure of Mo-graphene edge contacts is optimized manually. For this manual relaxation, we initially choose the Mo(110)-graphene-Mo(110) structure that is shorter than what we discuss in the main article. First, we prepare independent structures of a bulk Mo and a graphene sheet. Next, since the lattice mismatch between Mo(110) and the armchair edge of graphene is 4 %, we modify a lattice constant of Mo(110) to cancel out this mismatch. When we manually relax the structure, the Mo atoms are fixed at their initial position, and we adjust position of the graphene sheet. After the sequence of the manual relaxation, we remove Mo(110) on an edge, elongate the width of graphene ribbon, and passivate the Mo-removed edge with H.

The manual relaxation is performed as follows: (1) Firstly, we adjust spacing between Mo and graphene. The spacing with the minimum energy is 1.9 Å for this step. (2) Secondly, $z$ position of graphene is adjusted. We do not change the $z$ position because the initial position is already the optimized point. (3) Thirdly, $y$ position of graphene is adjusted while the spacing and $z$ position are fixed. To get the minimum energy, $y$ position is shifted by –0.71 Å. (4) Finally, spacing is adjusted again to 1.68 Å with the fixed $y$ and $z$ positions. Fig. S1(a)–(d) show details in this procedure. We also build Mo-graphene top contact structure by optimizing spacing between Mo surface and graphene sheet (Fig. S1(e)). For DFT-NEGF transport simulations, we modify the supercell structures optimized with the aforementioned method. Fig. S2 shows Mo-graphene edge and top contact geometries adopted as TranSIESTA inputs.

To build a structure for insertion of an O atom into the interface, we manually optimize the structure. Fig. S1(f) shows total energy as a function of the distance between the Mo surface and the O atom with respect to different $y$ positions. Since there could be a different point for optimization of the $y$ position, we test two different $y$ positions, $\Delta y$ = 0 Å and 1.575 Å. The structure with $\Delta y$ = 0 Å and spacing = 2.35 Å is the most stable (Fig. S1(f)), so that we have the optimized structure of the Mo-O-graphene edge contact shown in Fig. S1(d).

## 2. Density of states of Mo and graphene

Fig. S3 shows density of states for a bulk Mo and an infinite graphene sheet. When we make a contact of them, Mo $d$ orbital and carbon $p_z$ orbital, which is associated with graphene $\pi$ orbitals, would overlap with each other around the Fermi level. The Mo $d$ states mainly distribute within an energy range from –5 to 5 eV, so that electronic structures of the interface graphene are strongly modified within this range as shown in Fig. S3(a).

## 3. Width-dependent band gap of edge-contacted graphene nanoribbon

Different types of the armchair graphene nanoribbon (AGNR) exhibit distinct width dependences of band gap. It is known that, for H-passivated AGNRs, 3$n$+2 type AGNR shows the smallest band gap. As for Mo-contacted AGNRs, the 60 AGNR exhibits almost zero band gap whereas band gaps of 59 and 61 AGNRs are greater than 0.1 eV (Fig. S4). 59, 60, and 61 AGNRs are associated with $3n-1$, $3n$, and $3n+1$ types, respectively, where $n = 20$.

The band-decomposed charge density shows that the greatly reduced probability density of the $p_z$ orbitals at the interfacial carbon atoms for all those types. On the contrary, the probability density at the H-passivated carbon atoms on the opposite edge is large. As discussed in the main article, this could be account for the different behavior of the width-dependent band gap.

**4. Doping type of graphene for varying distance between Mo surface and O atom**

In the main article, we check the doping type of the interface graphene. By inserting an O atom, we can realize $n$-type graphene, which is consistent with experimental results. This results from the optimized distance of 2.35 Å between Mo surface and O atom. However, the spacing can spatially vary at the interface, so that we carry out additional calculations for the doping type with changing the spacing (Fig. S5). As a result, the $n$-type graphene is still obtained with respect to a wide range of spacing from 1.5 Å to 2.7 Å, total energies of which are shown in Fig. S1(f).

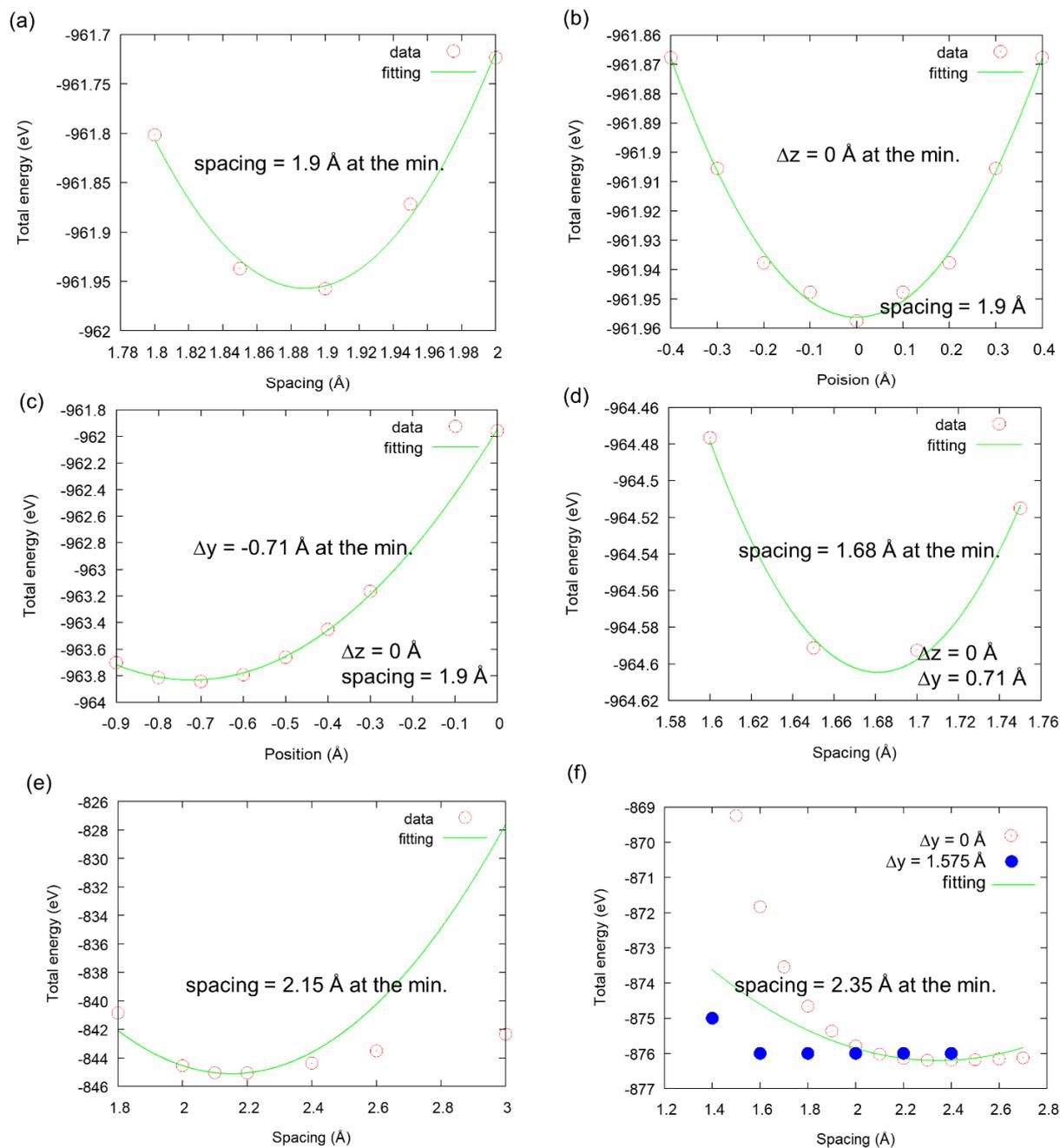

Fig. S1. Manual relaxation procedure for Mo-graphene edge contact structure: (a) Firstly, spacing is adjusted. (b) Secondly, *z* position of graphene is adjusted with the adjusted spacing. (c) Thirdly, *y* position of graphene is adjusted with obtained spacing and *z* position. (d) Finally, spacing is adjusted again with the adjusted *y* and *z* positions. (e) We also optimize Mo-graphene top contact structure by adjusting spacing between Mo surface and graphene sheet. (f) Total energies as functions of spacing for building the Mo-O-graphene structure.

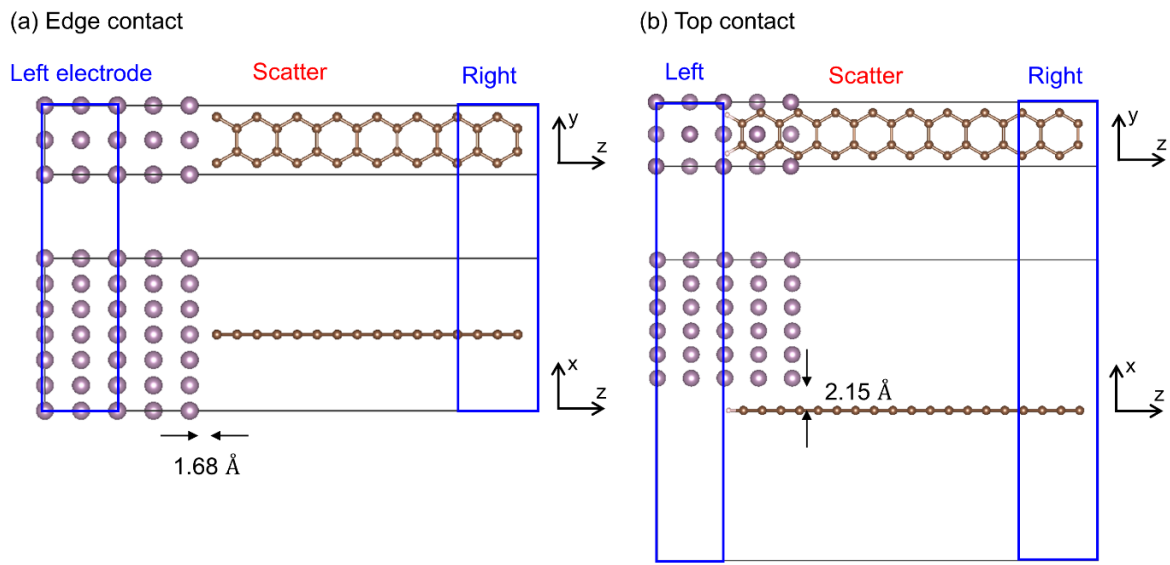

Fig. S2. Mo-graphene (a) edge and (b) top contact structures for DFT-NEGF transport simulations. Upper and lower figures correspond to top and side views, respectively.

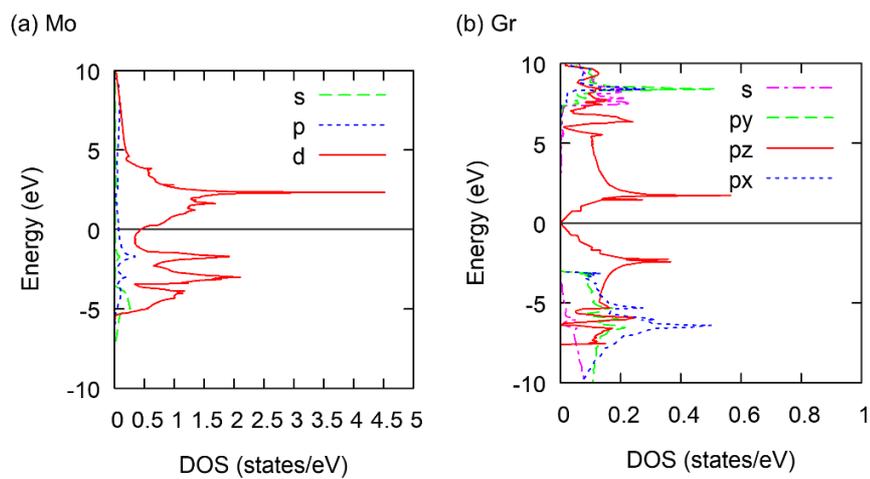

Fig. S3. Density of states for (a) a bulk Mo and (b) an infinite graphene sheet. The Fermi level is set to zero.

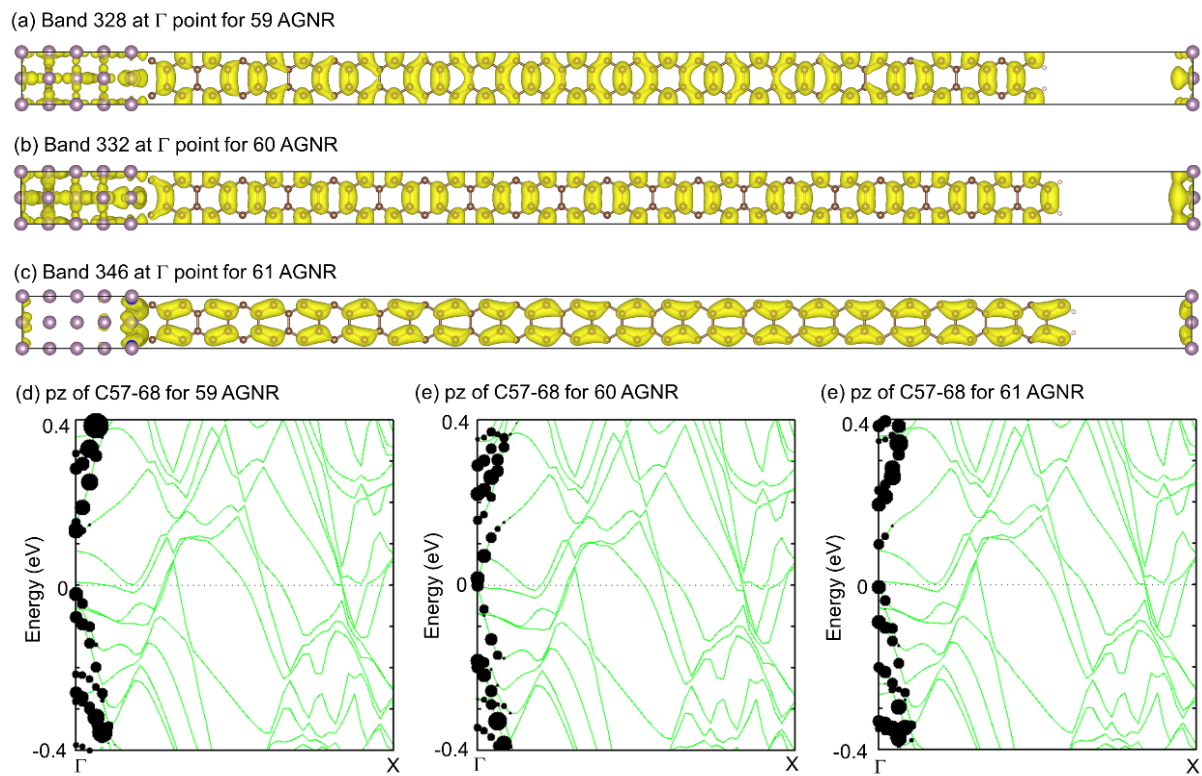

Fig. S4. Band-decomposed charge density for (a) 59, (b) 60, and (c) 61 AGNRs, which correspond to three distinguishing types, $3n-1$, $3n$ and $3n+1$. Orbital-decomposed band structures for (a) 59, (b) 60, and (c) 61 AGNRs. The Fermi level is set to zero.

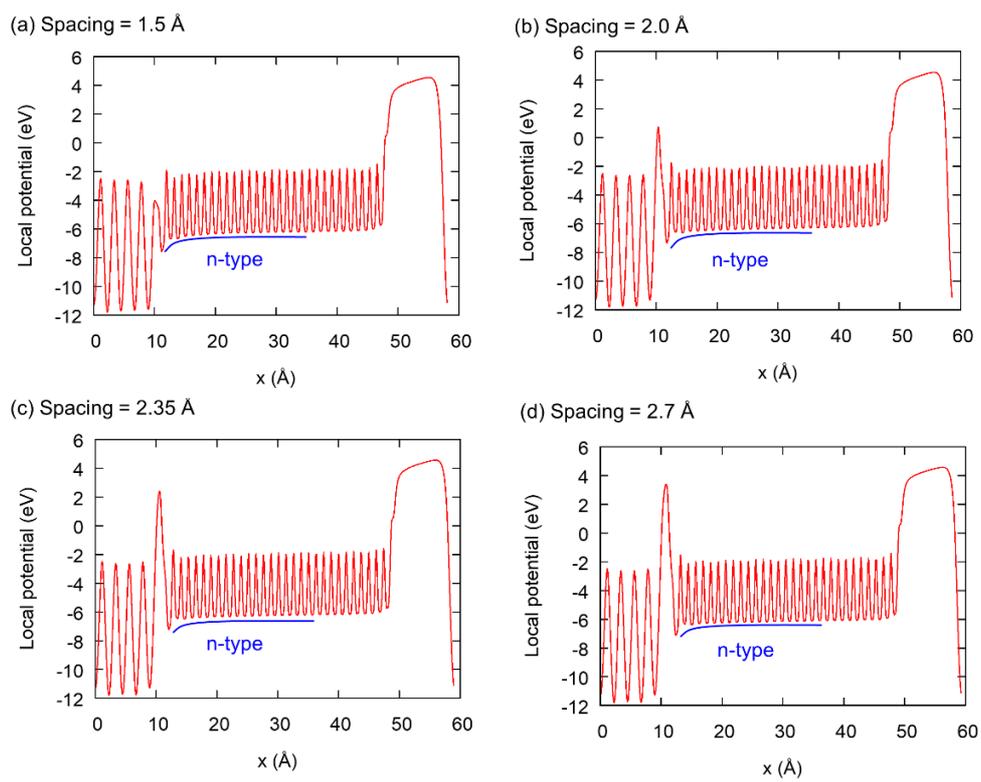

Fig. S5. *yz*-plane averaged local potential of Mo-O-graphene edge contact for different spacings: (a) 1.5 Å, (b) 2.0 Å, (c) 2.35 Å, and (d) 2.7 Å. The Fermi level is set to zero. The local potential here corresponds to a summation of the ionic and Hartree potentials. For a *yz*-plane average, a local potential is averaged over *z* within the interlayer distance of graphite, i.e., 3.4 Å.